\documentclass[%
 reprint,
 amsmath,amssymb,
 aps,
]{revtex4-2}

\usepackage{graphicx}%
\usepackage{dcolumn}%
\usepackage{bm}%
\usepackage[english]{babel}

\usepackage[utf8]{inputenc}
\usepackage{blindtext}
\usepackage{bbold}
\usepackage{amssymb}
\usepackage{amsmath}
\usepackage{commath}
\usepackage{braket}
\usepackage{siunitx}
\usepackage{graphicx}
\usepackage[]{algorithm2e}
\usepackage{hyperref}
\usepackage{nameref}
\usepackage{MnSymbol}
\usepackage{placeins}
\usepackage{todonotes}
\usepackage{glossaries}
\glsdisablehyper

\newacronym{rnn}{RNN}{recurrent neural network}
\newacronym{nn}{NN}{neural network}
\newacronym{lstm}{LSTM}{long short-term memory}
\newacronym{gsc}{GSC}{gate set calibration}
\newacronym{gst}{GST}{gate set tomography}
\newacronym{qip}{QIP}{quantum information processor}
\newacronym{qp}{QP}{quantum processor}
\newacronym{mae}{MAE}{mean absolute error}
\newacronym{mse}{MSE}{mean squared error}
\newacronym{rmse}{RMSE}{root mean squared error}
\newacronym{gru}{GRU}{gated recurrent unit}
\newacronym{dnp}{DNP}{dynamic nuclear polarization}
\newacronym{awg}{AWG}{arbitrary waveform generator}
\newacronym{dcg}{DCG}{dynamically corrected gates}
\newacronym{nisq}{NISQ}{noisy intermediate-scale quantum devices}
\newacronym{sgd}{SGD}{stochastic gradient descent}
\newacronym{rb}{RB}{randomized benchmarking}
\newacronym{cptp}{CPTP}{completely positive and trace preserving}

\begin{document}

\title{Data-Driven Qubit Characterization and Optimal Control using Deep Learning}

\author{Paul Surrey}
\email{paul.surrey@rwth-aachen.de}
\affiliation{JARA-FIT Institute for Quantum Information, Forschungszentrum Jülich GmbH and RWTH Aachen University, 52074 Aachen, Germany}

\author{Julian D. Teske}
\affiliation{JARA-FIT Institute for Quantum Information, Forschungszentrum Jülich GmbH and RWTH Aachen University, 52074 Aachen, Germany}

\author{Tobias Hangleiter}
\affiliation{JARA-FIT Institute for Quantum Information, Forschungszentrum Jülich GmbH and RWTH Aachen University, 52074 Aachen, Germany}

\author{Hendrik Bluhm}
\email{bluhm@physik.rwth-aachen.de}
\affiliation{JARA-FIT Institute for Quantum Information, Forschungszentrum Jülich GmbH and RWTH Aachen University, 52074 Aachen, Germany}

\author{Pascal Cerfontaine}
\email{pascal.cerfontaine@th-koeln.de}
\affiliation{Institute of Computer and Communication Technology, TH Köln, 50679 Köln, Germany}

\begin{abstract}

Quantum computing requires the optimization of control pulses to achieve high-fidelity quantum gates. We propose a machine learning-based protocol to address the challenges of evaluating gradients and modeling complex system dynamics. By training a \gls{rnn} to predict qubit behavior, our approach enables efficient gradient-based pulse optimization without the need for a detailed system model. First, we sample qubit dynamics using random control pulses with weak prior assumptions. We then train the \gls{rnn} on the system's observed responses, and use the trained model to optimize high-fidelity control pulses. We demonstrate the effectiveness of this approach through simulations on a single $ST_0$ qubit.

\end{abstract}
        
\maketitle

\section{Introduction}
\label{sec:Introduction}
\label{ssec:Solved_Problem}

\begin{figure*}[!t]
    \begin{center}
    \includegraphics[width=1\linewidth]{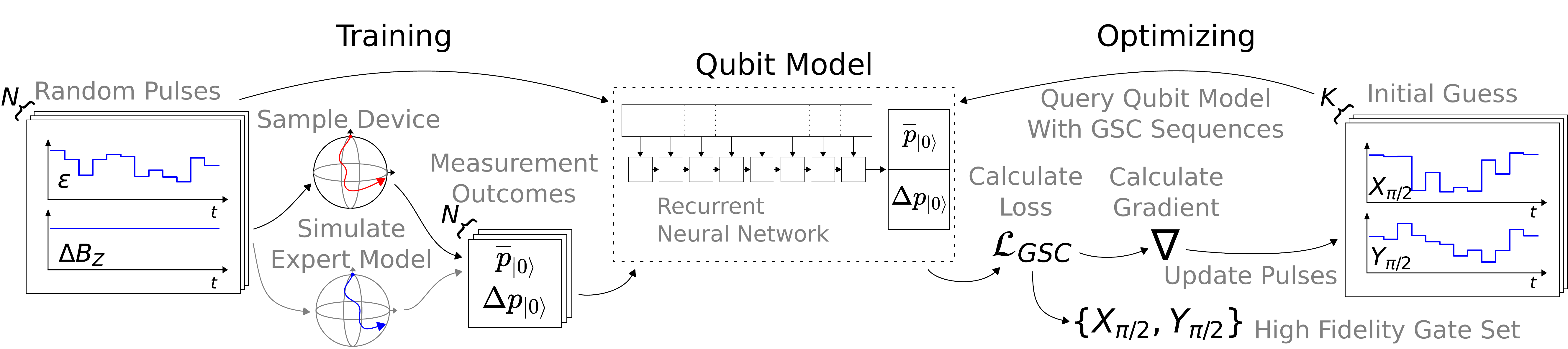}
    \caption{
        Diagram of our model training and pulse optimization pipeline.
        We model the qubit using a \gls{rnn}. 
        The network is trained on $N$ randomly chosen control pulses and the measurement outcome from the device or a simulation sampled with these pulses (left). 
        Once the network is trained, it is used as a proxy for the qubit to perform pulse optimization by means of \gls{gsc} \cite{Cerfontaine_2020} (right). 
        To this end, randomly initialized pulses are first concatenated to form sequences inspired by \gls{gsc}. For each sequence, the qubit model predicts measurement outcomes, from which the \gls{gsc} loss $\mathcal{L}_{GSC}$ is calculated.
        The pulse parameters are then updated in a gradient-based optimization scheme by minimizing $\mathcal{L}_{GSC}$.
    }
    \label{fig:protocol}
    \end{center}
\end{figure*} 
  
High-fidelity quantum gates are a necessary ingredient for scalable quantum computing \cite{DiVincenzo_2000} and require a precise understanding of the qubit's dynamics and careful tuning of the available control parameters \cite{Khaneja2005OptimalCO, PhysRevA.84.022305, Teske_2019, Glaser_2015, PhysRevLett.120.150401}. 
In practice, one is often faced with incomplete knowledge of how these commonly time-varying control parameters translate into the evolution of quantum systems \cite{Teske_2019, Cerfontaine_2020}, making the task of obtaining high-fidelity control pulses difficult. 
For instance, the dependence of the qubit's evolution on the physically applied control parameter, bandwidth limitations, transients of the control signals are typically not known with high enough precision as they depend on many aspects of the experimental setup \cite{Wittler_2021, PhysRevLett.110.146804, Cerfontaine_2014}.
This leads to model deficiencies, which limit the achievable gate fidelity in offline pulse optimization \cite{Cerfontaine_2020}.

Several techniques have been demonstrated to obtain and optimize control pulses to reach high-fidelities \cite{Cerfontaine:768380, Cerfontaine_2014, Cerfontaine_2020, Cerfontaine_2020_3, Dong_2015, PhysRevA.79.053417, Kelly_2014, Zahedinejad_2015, Zahedinejad_2016, Wittler_2021, Krastanov_2019, Khaneja2005OptimalCO, Dalgaard2020-fo, ernst2025reinforcementlearningquantumcontrol, david2024longdistancespinshuttling, genois2024quantumoptimalcontrolsuperconducting, ma2025machinelearningestimationcontrol, shindi2023modelfreequantumgatedesign}. 
Studies have shown that closed-loop optimization schemes incorporating direct experimental feedback tend to achieve higher fidelities \cite{Cerfontaine_2020, Kelly_2014}. 
In practice, pulses obtained from open-loop methods, which operate without direct experimental feedback \cite{Khaneja2005OptimalCO}, often serve as initial guesses for these closed-loop procedures \cite{Cerfontaine_2020}.  
However, open-loop approaches alone generally fail to reach high fidelities \cite{Cerfontaine_2020_3, Kelly_2014, Rol_2017}.
This discrepancy arises not from the absence of an intermediate model, but from inaccuracies and mismatches between the model used for pulse optimization and the true experimental system.
Efficient gradient computation is feasible only in simulation, whereas closed-loop optimization must rely on experimental feedback rather than analytical gradients. Consequently, hybrid strategies are commonly employed \cite{Cerfontaine_2020, Teske_2019, PhysRevApplied.17.024006}.
To not rely on some potentially deficient model for the open-loop optimization, one can use neural networks as universal function approximators \cite{HORNIK1989359} trained on real measurements to model the device \cite{Perrier_2020, Flurin_2020, genois2021quantumtailored, 8539575, August_2017, Teoh_2020, ZENG2020126886, PhysRevA.97.042324, Valenti_2019, PhysRevA.110.032607, PhysRevApplied.19.044090}. In this way, the system specifics are extracted from the provided data and not from theoretical descriptions. Including more knowledge about the quantum system into the \gls{nn}'s architectures could lead to more interpretable models, but as shown by \citet{genois2021quantumtailored} this does not guarantee higher accuracies.

To address the challenges of optimizing quantum control pulses, we explore the use of \glspl{rnn}, which are well-suited for modeling dynamic systems due to their ability to learn temporal dependencies from data. In this work, we tackle the problems of costly finite-difference gradient approximations, non-trivial initial guesses, and deficient prior models using a data-driven approach for qubit characterization and optimal control. We show through simulations that a \gls{rnn} can learn non-trivial system dynamics without extensive prior knowledge of the device, relying solely on the system's response to random control pulses. We then use the trained network as a surrogate model for the simulated physical system to perform gradient-based offline pulse optimization of a set of single-qubit gates, benefiting from the network's differentiability. Finally, we validate this protocol on two different device simulations.

\subsection{Previous Work}
\label{ssec:Previous_Work}
Several techniques exist to increase the precision of open-loop optimizations compared to closed-loop optimizations. They can be broadly divided into two categories: model-free and model-based estimation methods. Model-free approaches \cite{August_2017, Kelly_2014, Rol_2017, Zahedinejad_2015, Zahedinejad_2016, Teoh_2020} bypass explicit modeling and rely entirely on data-driven optimization, making these methods more general but also more challenging. In model-based methods, one tries to learn the parameters of a predefined model; this is usually the Hamiltonian of the physical system, ideally complemented with some noise model \cite{Krastanov_2019, Wittler_2021, Perrier_2020, Cerfontaine_2014}.

Using a Hamiltonian to model the dynamics of a \gls{qp} was demonstrated in STEADY \cite{Krastanov_2019}. However, this is insufficient for a complete description since \glspl{qp} typically interact with their environment.
Consequently, Lindblad-type master equations are often employed to extend Hamiltonian models with noise and decoherence terms \cite{Wittler_2021}, though they still rely on approximate assumptions about the underlying physical processes.
More general models can also be employed. For example, \citet{Flurin_2020, koolstra2021monitoring} use an \gls{rnn} to describe qubit dynamics, incorporating weak measurement feedback. \citet{genois2021quantumtailored} compare the accuracy of a Lindblad master equation based and an \gls{rnn}-based model in predicting the dynamics of a superconducting qubit coupled to a cavity. Their findings indicate that while the parameters of the master equation based model are more interpretable due to their alignment with the Lindblad stochastic master equation, the \gls{gru}-based \gls{rnn} achieves higher prediction accuracy when trained on large amounts of real experimental data.

Regardless of whether a model-based or model-free approach is used, pulse optimization typically requires a well-defined objective. The entanglement fidelity measure, for instance, can be used as a loss function to optimize control pulses \cite{qopt}. Measuring this fidelity experimentally requires advanced tomography methods, such as \gls{gst} \cite{nielsen2020gate, https://doi.org/10.48550/arxiv.1310.4492, Dehollain_2016, White_2021} or \gls{rb} \cite{Kelly_2014, Wallman_2014}, which are time-consuming. 

\subsection{Motivation and Approach}
\label{ssec:Motivation_and_Approach}

In this work, we build on STEADY \cite{Krastanov_2019} and $\mathbb{C}^3$ \cite{Wittler_2021}, which demonstrate that system responses can be learned from measurements of randomly applied control pulses. We also incorporate insights from \cite{Flurin_2020, koolstra2021monitoring, genois2021quantumtailored}, where an \gls{rnn} is used to model time-dependent qubit dynamics. By synthesizing these approaches, we propose a fully data-driven method for qubit characterization and optimal control.

Encoding the dynamics of a quantum system using a \gls{nn}, a universal function approximator \cite{HORNIK1989359}, avoids the biases and constraints of incomplete analytical models, instead shifting the modeling challenge to the choice of the \gls{nn} architecture. 
The quality of this data-driven model depends on the sampling procedure, the number of measured data points, and the design of the neural network, including its architecture and size. While purely data-driven methods generally require significantly more data to achieve the same accuracy as analytical models, acquiring additional data may be more practical or cost-effective in typical quantum computing experiments, where thousands or even millions of data points can be collected per second.

Leveraging this experimental scalability, we sample the device using random control pulses, henceforth referred to as probe pulses. This approach follows \cite{Krastanov_2019}, where some random pulses were chosen to incorporate commonly available and inexpensive system knowledge. 
A \gls{rnn} is then trained to predict the experimentally accessible measurement outcomes from these system samples, serving as a surrogate model for the device or simulation. This trained model is subsequently used to optimize control pulses via gradient descent. Our approach is outlined in more detail in Fig. \ref{fig:protocol}. 

The loss function for the pulse optimization is obtained from the \glsfirst{gsc} protocol \cite{Cerfontaine_2020} because of its simplicity, self-consistency, and its experimentally accessibility. 
Alternatively, our surrogate model could also be used to optimize control pulses with protocols such as \gls{gst} \cite{nielsen2020gate} and \gls{rb} \cite{Kelly_2014}.
Using the \gls{gsc} loss results in a differentiable loss function and serves as a proxy for the gate infidelity, making efficient use of the surrogate models predictions.

A decisive advantage of using an end-to-end \gls{nn} to model the full dynamics of a \gls{qip} is the ability to leverage the extensive ecosystem of \glspl{nn}. These models are easy to evaluate, exchange, and share, enabling researchers to test their algorithms across different devices in simulation. Moreover, modern \gls{nn} frameworks are highly optimized for efficient computation, potentially allowing \gls{nn}-based models to be evaluated faster than classical simulations relying on analytical models. The gradient calculations through the \gls{nn} benefits from backpropagation, enabling efficient first-order gradient calculations.

Our paper is structured as follows:  
We begin by introducing the notation used throughout the work in Sec.~\ref{sec:General_Qubit_Realizations}.  
Next, we describe the sampling approach in greater detail and provide a detailed discussion of how the \gls{gsc} protocol is incorporated in Sec.~\ref{sec:GSC}.  
We then outline the reasoning behind our choice of \gls{nn} architecture and optimization strategy in Sec.~\ref{sec:Neural_Network}.  
Finally, in Secs.~\ref{sec:Pulse_Optimization} and \ref{sec:Specific_Realization}, we demonstrate our method on two different \glspl{qip}, analyzing both the learned \gls{qip} model and the resulting gate performance.

\section{Notation}
\label{sec:General_Qubit_Realizations}
                
In the following, we introduce the model and notation for a closed quantum system. Our method does not depend on a specific qubit Hamiltonian, but rather learns the relationship between control inputs and projective measurement outcomes, making it broadly applicable and qubit-agnostic.
                
Dynamics of a qubit are typically described using a Hamiltonian (with $\hbar = 1$),
\begin{equation}
        \hat{H}(t, \boldsymbol{\lambda}, \boldsymbol{\delta \lambda})
        = \hat{H}\textsubscript{c}(t ; \boldsymbol{\lambda})
        + \hat{H}\textsubscript{n}(t ; \boldsymbol{\lambda}, \boldsymbol{\delta \lambda})
        \label{eq:main_H}
\end{equation}
where $\hat{H}\textsubscript{c}(t ; \boldsymbol{\lambda})$ describes the qubit evolution under the control input tensor $\boldsymbol{\lambda}$. The elements of $\boldsymbol{\lambda}$ are the control amplitudes $\lambda_{j, t}$ applied to parameter $j$ from time $t$ to $t+\Delta t$. 
$\hat{H}\textsubscript{n}(t ; \boldsymbol{\lambda}, \boldsymbol{\delta \lambda})$ captures the contribution of noise to the system dynamics, based on a specific noise realization $\boldsymbol{\delta \lambda}$, where the subscript $n = 1, \dots, N$ indexes the $N$ Monte Carlo noise samples.
For simplicity, we assume a fixed time step of $\Delta t =  \SI{1}{\nano\second}$, such that the control pulse is defined as a sequence of discrete segments. The control Hamiltonian at time step $t$ is given by
\begin{equation}
    \hat{H}\textsubscript{c}(t ; \boldsymbol{\lambda}) = \sum_j \lambda_{j, t} \hat{\sigma}_j
    \label{eq:H_c}
\end{equation}
where $\lambda_{j, t}$ denotes the control amplitude applied to operator $\hat{\sigma}_j$ at time $t$. Note that these amplitudes may not be linearly related to the physically applied control fields.

The system's response to a sequence of control fields, i.e., a control pulse, when initialized in $\ket{0}$ and measured in the same state after evolution, is given by
\begin{equation}
    \overline{p}_{\ket{0}}(\boldsymbol{\lambda}) = \frac{1}{N} \sum_n |\braket{0|\hat{U}_n(\boldsymbol{\lambda})|0}|^2,
\end{equation}
where the time evolution operator is given by the product of operators at each time step, with the full evolution from $t=0$ to $t=T-1$ represented as
\begin{equation}
    \hat{U}_n(\boldsymbol{\lambda}) = \hat{U}_n{(\boldsymbol{\lambda}, t=T-1)} ... \hat{U}_n{(\boldsymbol{\lambda}, t=0)},
\end{equation}
with
\begin{equation}
    \hat{U}_n(\boldsymbol{\lambda}, t) = \exp(-i
    \hat{H}(t, \boldsymbol{\lambda}, \boldsymbol{\delta \lambda}_n)
    \Delta t).
\end{equation}
$T$ is the total duration of a pulse. The number of control amplitudes $L$ is then given by $ T = L \cdot \Delta t$. The system evolves under the control input $\boldsymbol{\lambda}$ and a noise realization $\boldsymbol{\delta\lambda}_n$, resulting in the unitary evolution $\hat{U}_n(\boldsymbol{\lambda}, \boldsymbol{\delta\lambda}_n)$.

To quantify the noise-induced stochastic fluctuations, we compute the standard error $\Delta p_{\ket{0}}(\boldsymbol{\lambda})$ across $N$ Monte Carlo samples, defined as
\begin{equation}
    \Delta p_{\ket{0}}(\boldsymbol{\lambda}) = \frac{1}{N} \sqrt{\sum_n (|\braket{0|\hat{U}_n(\boldsymbol{\lambda})|0}|^2-\overline{p}_{\ket{0}}(\boldsymbol{\lambda}))^2}.
\end{equation}
            
\section{Gate Set Calibration}
\label{sec:GSC}

The \gls{gsc} protocol addresses systematic errors in gate sets on incompletely characterized systems. Applied in closed-loop with real-world \gls{qip} devices, it has enabled self-consistent, high-fidelity gate sets for semiconductor-based spin qubits \cite{Cerfontaine_2020}.

In an open-loop optimization setting, the reached gate fidelity is typically limited by imperfect system tuning, incomplete knowledge of the coherent device dynamics (e.g., transfer functions, nonlinear parameter dependencies), and by decoherence processes such as noise-induced relaxation and dephasing. These uncertainties cause control pulses optimized in simulation to deviate from their intended behavior on hardware. Moreover, fidelity measures can themselves be biased by systematic errors in the measurement process, leading to inaccurate optimization outcomes.

\gls{gsc} mitigates these issues by identifying and correcting systematic errors, modeling them as unintended Pauli operations acting alongside the intended gates. While not entirely insensitive to SPAM (state preparation and measurement) errors, \gls{gsc} is designed such that their effect is weak enough to not compromise achievable fidelities.

After selecting a gate set, the gates are arranged into \gls{gsc}-sequences. Evolving an initial qubit state under these sequences and performing a projective measurement afterward yields measurement outcomes that depend to first order on the individual gate errors for suitably chosen sequences.

The differences between the expected outcomes and the measured outcomes of the chosen sequences yields the so-called error syndrome. Even without solving for the individual Pauli errors, the syndrome is a proxy for the coherent gate infidelities, i.e. the systematic gate errors. 
        
In this work we use the concept of \gls{gsc} to obtain the loss function 
\begin{equation}
    \mathcal{L}_{GSC} = 
    \sum_i \frac{1}{N_{seq}}
    \left|
    {
    \Delta 
    \mathcal{R}_i
    }
    \right|^2.
    \label{eq:L_GSC}
\end{equation}
where 
\begin{equation}
    \Delta \mathcal{R}_i = \mathcal{R}(\lambda_i) - R_i^0
\end{equation}
is the error syndrome component. $\mathcal{R}(\lambda_i)$ is the measured outcome obtained from applying pulse $\lambda_i$ in syndrome sequence $i$, and $R_i^0$ is the corresponding expected (ideal) outcome.
This loss function is minimized when optimizing the pulses using the trained \gls{nn}. Further details on \gls{gsc} can be found in \cite{Cerfontaine_2020, AragonesSoria2022}.

The simulations presented in the \gls{gsc} publication \cite{Cerfontaine_2020} show that initial guesses must have an infidelity of at most $I \lesssim 20\%$ to converge on a $ST_0$-qubit device \cite{Cerfontaine_2020, Cerfontaine_2014}.

\section{Neural Network}
\label{sec:Neural_Network}
As mentioned in the introduction, we train a \gls{nn} to represent the functional relation between the input pulse and the measurement outcome.
Common knowledge about typical experimental setups, qubit dynamics, the non-commutativity of Hamiltonian operators, and signal distortions (often modeled as causal filters) suggests that \glspl{rnn} are a suitable network architecture. This introduces an inductive bias, i.e. a preference for functions with temporal dependencies and local correlation, but with a sufficient number of layers and parameters the neural network acts as an arbitrary function approximator, which is less constraining than incomplete models \cite{genois2021quantumtailored}.

On one hand, parts of the control problem, like bandwidth limitations, do not depend on a specific position in time.
On the other hand, the operations performed on the qubit do not generally commute and therefore their time-ordering matters.
These symmetries of the control problem can be exploited by using corresponding network architectures.
Hence we use convolutions to represent time-independent features and a recurrent layer for time-dependent features. 
For the recurrent layer, we use a standard \gls{lstm} \cite{HochSchm97} implementation. 
The last hidden state is reduced to 2 dimensions (i.e., $ \overline{p}_{\ket{0}}(\boldsymbol{\lambda}) $ and $ \Delta p_{\ket{0}}(\boldsymbol{\lambda}) $) using standard dense layers. 
The exact architecture used for our experiments is given in the appendix (Tab.~\ref{tab:nn:architecture_summary}).

While training the \gls{nn}, we do not explicitly enforce physicality. 
We do not choose the shape of the \gls{nn} to precisely fit common theoretical models, nor do we penalize missing \gls{cptp} properties of intermediate representations that the model might have learned. We likewise refrain from embedding elements of established analytical models into the functional relationship between $\boldsymbol{\lambda}$ and the outputs $\{ \overline{p}_{\ket{0}}, \Delta p_{\ket{0}} \}$.

Alongside the common error metrics \gls{mae} and \gls{rmse}, we also evaluate network performance using the accuracy measure $\mathcal{A}_d$ (Eq. \ref{eq:accuracy_definition}), defined as the percentage of predictions with a smaller absolute error than some fixed bound.
$\mathcal{A}_{d}$ is defined as follows:
\begin{equation}
    \mathcal{A}_{d}
    = \frac{1}{N} \sum_i{
            \begin{cases}
                1 & \text{for }\abs{t_i - f(x_i)} \leq d \\    
                0 & \text{else}
            \end{cases} 
        }
    \label{eq:accuracy_definition}
\end{equation}
where $t_i$ is measurement $i$ out of $N$ and $f(x_i)$ is the network's prediction for this data point.

The networks used in Ref.~\cite{genois2021quantumtailored} include additional losses to constrain the output and continuity of the network predictions to physical values. In our work, we also constrain the output of the network by clipping the output value to the physical range (i.e. in $[0, 1]$). We train our network on pulses with different pulse lengths ($L \in [10, 50] \text{ segments}$). Therefore, information of the total length of the pulse is only known for the last $m$ pulse segments. With the chosen network architecture $m=3$.

The network is trained using Adam \cite{kingma2017adam}, using \gls{mae} as the loss function for the training, and sample weighting for an even distribution over $\overline{p}_{\ket{0}}$. 
Further details can be found in the appendix (Sec.~\ref{sec:nn:architecture_summary}).

\begin{figure}[]
    \includegraphics[width=0.95\linewidth]{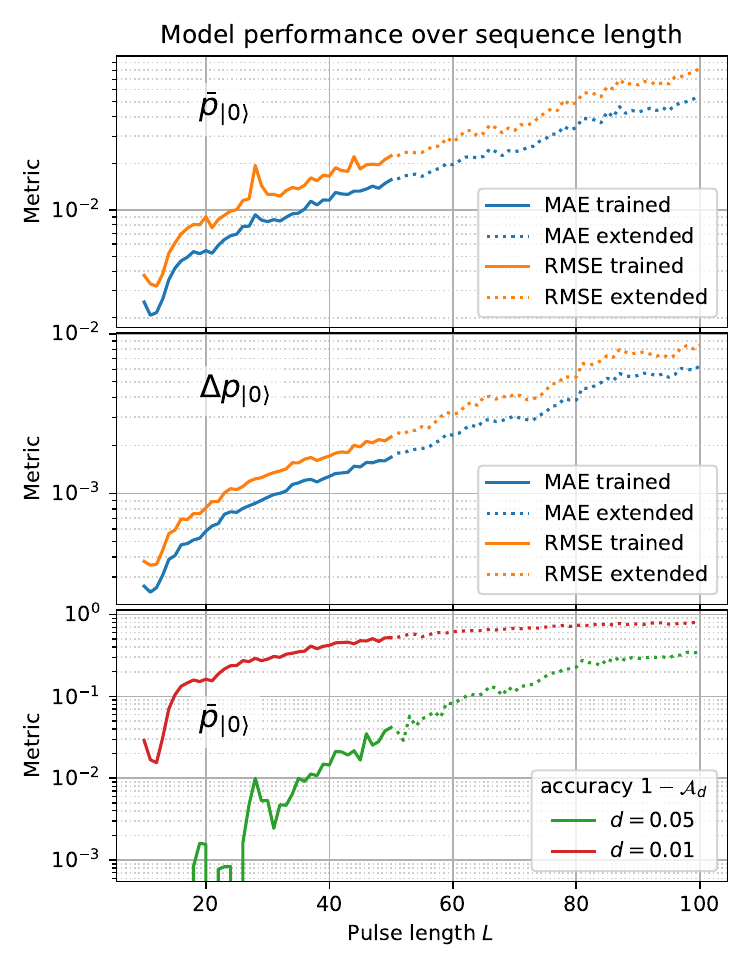}
    \caption{
        Network error for different control sequence lengths measured on the test set. The network is only trained on lengths corresponding to the solid line. Pulses of lengths corresponding to dotted lines have not been seen in the training phase and are not used for optimizing pulses, as the syndromes only require pulses of length $L<=50$.
    }
    \label{plot:nn:simple:performance_vs_length}
\end{figure}

We observed the model prediction error to only increase moderately for pulse lengths outside of the $L \in [10, 50]$ window as discussed in Fig.~\ref{plot:nn:simple:performance_vs_length}.

We observe that the model performs reasonably well on sequence lengths outside its training range. For both $\overline{p}_{\ket{0}}$ and $\Delta p_{\ket{0}}$, the trends in Fig.~\ref{plot:nn:simple:performance_vs_length} remain consistent between the trained lengths and the extended region. Nonetheless, the prediction accuracy gradually decreases as the model is pushed farther beyond the properties it was trained on.

\section{Pulse Optimization}
\label{sec:Pulse_Optimization}
    Backpropagation enables us to efficiently compute the gradient of the differentiable fidelity proxy $\mathcal{L}_{GSC}$ with respect to the voltage values of the control pulse. We use this gradient to optimize the pulses via stochastic gradient descent by directly minimizing $\mathcal{L}_{GSC}$. To further guide the optimization, we add an $L^2$-penalty on the predicted standard deviation of the measurement outcomes for the \gls{gsc} sequences, encouraging the algorithm to suppress decoherence-related errors. Finally, we include a small regularization term that averages gradients across each batch of pulses, which helps stabilize and smooth the optimization dynamics. The full update procedure for the control-pulse voltages is provided in Algorithm~\ref{alg:opt:general_secription} in the appendix.

    We run the optimization for $K$ gate sets on the same model in parallel. These optimization runs are initialized with different random control values. 
    However, for the purpose of initializing \gls{gsc} for closed-loop optimization on an experiment only a handful of the pulses surpassing the initial fidelity threshold will be necessary.
    Thus, we reduce the number of pulse sets by choosing the 10 out of 256 gate sets with the smallest $\mathcal{L}_{GSC}$ loss after the optimization. 
    This criterion is limited, as $\mathcal{L}_{GSC}$ only estimates systematic errors. If this is limiting, \gls{rb} could also be used to improve the selection.
    Calculating the average entanglement infidelity \cite{qopt} of the obtained pulses can be used as a quantitative measure of how well the learned model and pulse optimization work.
    To assess the performance of the pulse optimization we also report the relative number of gate sets reaching infidelities below $\leq 1\%$ (see Sec.~\ref{ssec:Pulse_Optimization}).

\section{Protocol Validation}
\label{sec:Specific_Realization}

    This section details the qubit platform, system simulations, surrogate-model training, pulse optimization procedure, and final characterization used to validate the proposed protocol. We focus on electron spin qubits, which naturally expose two impactful and experimentally relevant unknowns that challenge pulse optimization: hardware-induced transfer functions that distort control signals, and nonlinear exchange interactions that set the qubit rotation rate.

    To test the robustness of our approach under these uncertainties, we use measured system properties from \cite{PhysRevLett.110.146804} and \cite{Cerfontaine_2020_3} to define two different system simulations: a \textit{general} model for algorithm development and a more realistic \textit{specific} model for validation. The \textit{general} model is based on common approximations of the transfer function and the functional dependence of the exchange interaction on experimentally accessible control parameters. This model is primarily used for developing the algorithm. The \textit{specific} model incorporates experimentally measured transfer functions \cite{Cerfontaine_2020_3} and exchange interactions \cite{PhysRevLett.110.146804}.

\subsection{Qubit System and Model}
\label{ssec:Qubit_System_and_Model}

    In this work we consider a single $ST_0$ spin qubit implemented in two gate-defined quantum dots, each occupied by one electron.
    The singlet and one triplet state of the combined wave function of the two electrons form the qubit basis: $\ket{0} = \ket{S}$ and $\ket{1} = \ket{T_0}$. 
    The qubit rotation can be controlled by manipulating the exchange interaction $J(\vec{\epsilon}, t)$ between the two electrons \cite{PhysRevLett.89.147902}. $J(\vec{\epsilon}, t)$ is controlled by a voltage $\vec{\epsilon}$ which shifts the potential of one electron against the other (detuning). A second rotation axis of the qubit is controlled by the magnetic field gradient $\Delta B_Z$ between the two dots, which can be set to a constant value using \gls{dnp} \cite{PhysRevLett.104.226807} in GaAs or with micromagnets \cite{K_nne_2024} in Si-based systems . $\Delta B_Z$ can typically not be quickly changed during execution of a quantum circuit.

    In the $ST_0$ subspace, the Hamiltonian describing the dynamics of the singlet-triplet spin qubit can be written as
    \begin{equation}
        \hat{H}(t, \vec{\epsilon}, \Delta B_Z) = \frac{J(\vec{\epsilon}, t)}{2} \sigma_Z + \frac{\Delta B_Z}{2} \sigma_X.
    \end{equation}
    Here, 
    \begin{equation}
        J(\vec{\epsilon}, t) = J_0 \exp{\left(\frac{\epsilon^\prime_t}{\epsilon_0}\right)}
        \label{eq:exchange_interaction}
    \end{equation}
    is a fit to experimental data to describe the exchange interaction and 
    \begin{equation}
        \epsilon^{\prime}_t = \sum_\tau k(t-\tau)\epsilon_\tau
    \end{equation}
    is a linear transfer function capturing pulse distortions, where $k(t-\tau)$ describes the kernel (see Fig. \ref{plot:tf_kernel}).
    This notation fits Eq. \ref{eq:H_c} with $\lambda_{0, t} = J(\vec{\epsilon}, t)$, $\lambda_{1, t} = \Delta B_Z$ and $\vec{\sigma} = (\sigma_Z, \sigma_X)$.
    The magnetic field gradient is assumed to be set to $\Delta B_Z = 42.1 \cdot \si{2\pi 10^6 \frac{1}{\second}}$ using DNP \cite{Cerfontaine_2020_3}.

\subsubsection{Exchange Interaction}
\label{sssec:Exchange_Interaction}

    \begin{figure}[t]
        \includegraphics[width=\linewidth]{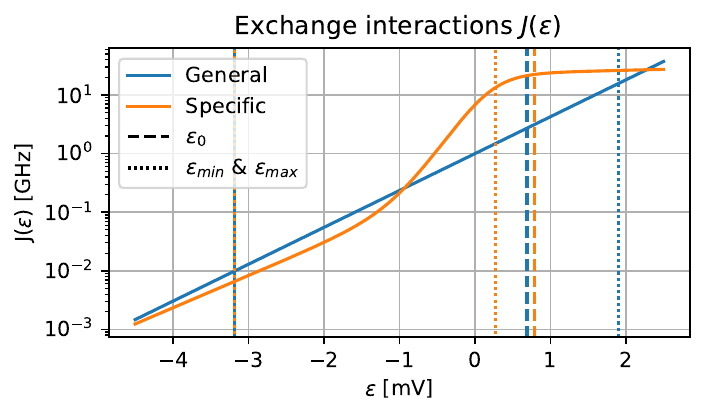}
        \caption{
        The exchange interactions used for simulations of the approximative \textit{general} model and the more accurate \textit{specific} model. $\epsilon_{min}$ and $\epsilon_{max}$ describe the voltage range in which random pulses are sampled. The same $\epsilon_{min}$ is chosen for both models. $\epsilon_0$ refers to the parameter in the exchange interaction (Eq. \ref{eq:exchange_interaction}).
        }
        \label{plot:exchange_interactions}
    \end{figure}

    The measurement by \citet{PhysRevLett.110.146804} shows that the exchange interaction can only be approximated in a certain voltage range by an exponential description $ J(\epsilon) = J_0 \exp(\epsilon/\epsilon_0) $ ($J_0 = 159.0 \cdot \si{2\pi 10^6 \frac{1}{\second}}$, $\epsilon_0 = \SI{0.69}{\milli\volt}$). 
    This approximation is implemented in the \textit{general} model.
    To show that our method also works for different systems, we used the measured exchange interaction (Fig. \ref{plot:exchange_interactions}) for the \textit{specific} simulations with a voltage range reaching into the regime of large divergence between the purely exponential dependence, and the measured exchange interaction.
    
\subsubsection{Transfer Function}
\label{sssec:Transfer_Function}
\begin{figure}[t]
        \includegraphics[width=1\linewidth]{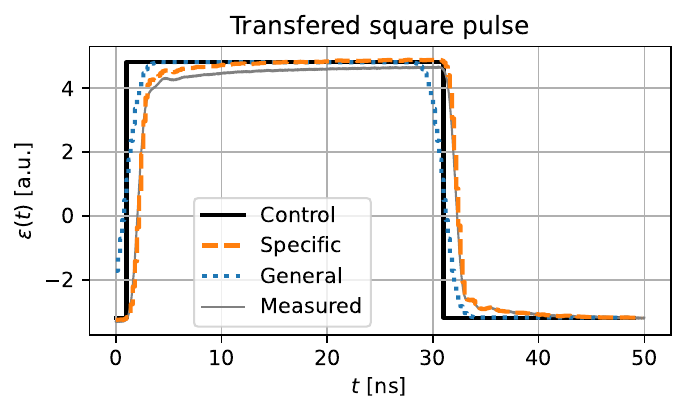}
        \caption{
            The transfer functions in relation to the programmed pulse (control) and the actually measured signal (measured). The qubit model using rough approximations is depicted as the dotted blue line (\textit{general}). The qubit model based on more realistic measurements is shown in dashed orange (\textit{specific}).
        }
        \label{plot:tf_kernel_step_function}
    \end{figure}

    The time resolution of the pulses needed for controlling GaAs singlet-triplet qubits lies in the \si{\nano\second} range, in which pulse distortions from the experimental setup are relevant (see Fig. \ref{plot:tf_kernel_step_function}). 
    The \glspl{awg}, used to convert the digital representation of the $\mathcal{O}(\si{\nano\second})$ time scale pulse into real voltages, and the connecting wires typically exhibit limited bandwidth and introduce transients.

\begin{figure}[t]
        \includegraphics[width=\linewidth]{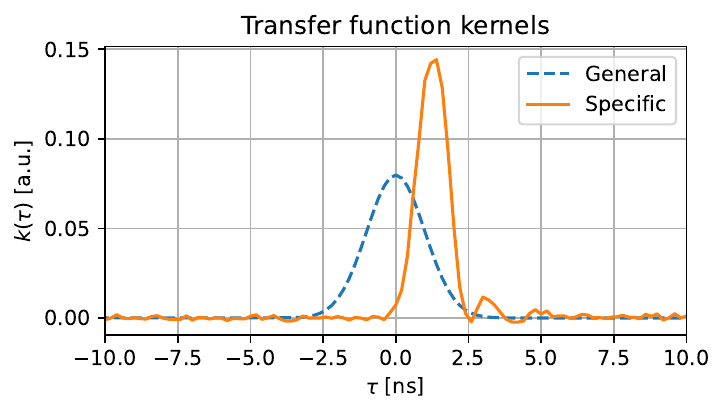}
        \caption{
            The approximated transfer function kernels used to model pulse distortions introduced by the signaling hardware. The transfer function in the \textit{general} simulation is an idealized, symmetric Gaussian kernel, while the transfer function in the \textit{specific} simulation is based on the measured system response and captures only the behavior after the \glspl{awg} begin outputting the corresponding pulse segment.
        }
        \label{plot:tf_kernel}
    \end{figure}

    In the \textit{general} model, the bandwidth limitation is represented by a Gauss kernel of width $\sigma = \SI{1}{\nano\second}$ (dashed blue line in Fig. \ref{plot:tf_kernel}) which is symmetric in time. Pulses filtered with this kernel kernel exhibit edge artefacts, visible at the start and end of each pulse (see the initial points of the dotted blue line compared to the black line in Fig. \ref{plot:tf_kernel_step_function}). 
    For the \textit{specific} model, we analyzed the measurements by \citet{Cerfontaine_2020_3} of the signal distortions across a standard coaxial cable and obtained a more realistic kernel (solid orange line in Fig. \ref{plot:tf_kernel}). 
    This kernel also contains transients and oscillations (The oscillations are on the order of $ < \SI{5}{\nano\second}$).

\subsubsection{Noise}
\label{sssec:noise}

    For investigations into creating pulses that are robust to quasi-static noise, i.e., \gls{dcg}, we implemented commonly used noise models: 
    Quasi-static noise drawn from a normal distribution acting on $\epsilon$ and $\Delta B_Z$, and fast noise added to $\epsilon$.
    Both the \textit{general} and the \textit{specific} model simulate the quasi-static noise with a standard deviation of $\sigma_{\epsilon} = \SI{8.0e-3}{\milli\volt}$ \cite{PhysRevLett.110.146804} and 
    $\sigma_{\Delta B_Z} = 2.8 \cdot \si{2\pi 10^6 \frac{1}{\second}}$ \cite{Cerfontaine_2020_3}. For fast noise both models use
    \begin{equation}
        S_\epsilon (f) =
        \begin{cases}
                \frac{S_0}{f^{0.7}} \cdot \SI{1}{\volt\squared\per\hertz}  & \text{for } \SI{5e4}{\hertz} \leq f \leq \SI{e6}{\hertz} \\
                \frac{S_0}{(10^6)^{0.7}} \cdot \SI{1}{\volt\squared\per\hertz} & \text{for } \SI{e6}{\hertz}   <  f \leq \SI{e10}{\hertz} \\
                 \SI{0}{\volt\squared\per\hertz} & \text{else}
        \end{cases},
        \label{eq:noise_model}
    \end{equation}
    where $S_0 = \SI{10.24e-16}{\volt\squared\per\hertz}$ \cite{PhysRevLett.110.146804}. Throughout this work the number of noise samples $N$ has been set to $N = 60$.
    The noise model of the \textit{specific} simulation only differs in $\Delta B_Z$ and $\sigma_{\Delta B_Z}$, where we added a small error the assumed $\Delta B_Z = 42.096 \cdot \si{2\pi 10^6 \frac{1}{\second}}$ and $\sigma_{\Delta B_Z} = 2.801 \cdot \si{2\pi 10^6 \frac{1}{\second}}$ used for the generation of the control sequences used to probe the device.
    For the simulation of the outcome of the probe pulses, the magnetic field gradient $\Delta B_Z$ is chosen to be constant for one pulse and is sampled from a normal distribution with $\mu = \Delta B_Z$ and $\sigma = \sigma_{\Delta B_Z}$. 
    The noise models are only used in the simulation of the outcomes of the probe pulses. Knowledge about the noise models is not used as prior knowledge in the device probing step and also not in the pulse optimization steps. 
    For comparing the fluctuations introduced by the noise model on the pulses, we calculate $\Delta p_{\ket{0}}$ with noise acting on the system and without. This noise model used in the \textit{general} device simulation introduces an average standard deviation of $\Delta p_{\ket{0}} = 3.7 \cdot 10^{-3}$ measured on $\SI{10}{\nano\second}$ long pulses simulated using the Schrödinger equation with and without noise. 
            
\subsection{Device Probing}
\label{ssec:Device Probing}
    
    We probe the device with minimal prior knowledge, using the exchange interaction from the \textit{general} model for a few probe pulses. This mitigates bias from the exponential dependence of the exchange interaction, which could otherwise distort the distribution. As the exchange interaction can be measured and approximated beforehand (Eq.~\ref{eq:exchange_interaction}), we incorporate this coarse knowledge but deliberately exclude assumptions about the transfer function.
    
    Concretely, a subset of the probe pulses is chosen with random voltages for each pulse segment in the range $[\epsilon_{min}, \epsilon_{max}]$.
    In addition, we sample another subset such that voltages lie in the region where $J(\epsilon) \cdot T \in [\pi/2, 4\pi]$ (this amounts to $10\%$ of all samples) and such that the angles between the $\ket{0}$ and $\ket{+}$ directions on the Bloch sphere, controlled by $J(\epsilon) \propto \exp(\epsilon)$, are sampled uniformly ($30\%$ of all samples). This method introduces pulses with longer sections of constant voltages for random durations.
    
    Focussing on one qubit, we work with the gate set $\mathcal{G}=\{X_{\pi/2}, Y_{\pi/2}\}$ where we probe $\mathcal{L}_{GSC}$ with sequences of all permutations of up to $4$ gates concatenated, including repetitions.
    We initialize the qubit before each sequence to $\ket{0}$ and measure in the Z-basis.
    A Z rotation of both gates from the gate set cannot be measured with the $\{X_{\pi/2}, Y_{\pi/2}\}$ gate set.
    Additionally sign changes will not be detectable, such that gates sets
    $\{X_{\pm\pi/2}, Y_{\pm\pi/2}\}$, 
    $\{Z_{-\theta} \cdot X_{\pi/2} \cdot Z_{\theta}, Z_{-\theta} \cdot Y_{\pi/2} \cdot Z_{\theta}\}$, and 
    $\{Y_{\pi/2}, X_{\pi/2}\}$ 
    will not be distinguishable via the used syndromes and hence $\mathcal{L}_{GSC}$. 
    (We discuss the calculations used to measure the infidelity in section \ref{ssec:appendix_gloabl_z_rotaion_consideration} of the appendix.)
    
    To predict the measurement outcomes of various gate sequences, the models must process control pulses with durations up to $T\textsubscript{total} \geq T$, where $T$ represents the duration of a single gate, and $T\textsubscript{total}$ is the longest sequence length considered. To achieve this, we measure $\overline{p}_{\ket{0}}(\boldsymbol{\lambda})$ and $\Delta p_{\ket{0}}(\boldsymbol{\lambda})$ for randomly sampled control sequences $\boldsymbol{\lambda}$ of varying lengths, up to  $T\textsubscript{total}$. 

\subsection{Trained Surrogate $ST_0$-qubit Model}
\label{ssec:Neural_Network}
    In this section we discuss the neural network architecture decisions and examine the neural networks performance.

\subsubsection{Network Architecture}
\label{sssec:Network_Architecture}
    The neural network uses a combination of convolutional layers (for local time-based features), an LSTM layer (to capture time-dependent dynamics), and dense layers (for the final prediction). The input is a time-series tensor representing control parameters, padded to a fixed maximum pulse length.
    The network outputs two predicted values (mean and fluctuation of measurement probabilities) constrained to valid probability ranges between $0$ and $1$. (A more extensive description of the model architecture can be found in the appendix in Sec.~\ref{sec:nn:architecture_summary}.)

\subsubsection{Characterization Results}
\label{sssec:Characterization_Results}
    
\begin{figure*}[]
        \includegraphics[width=0.9\linewidth]{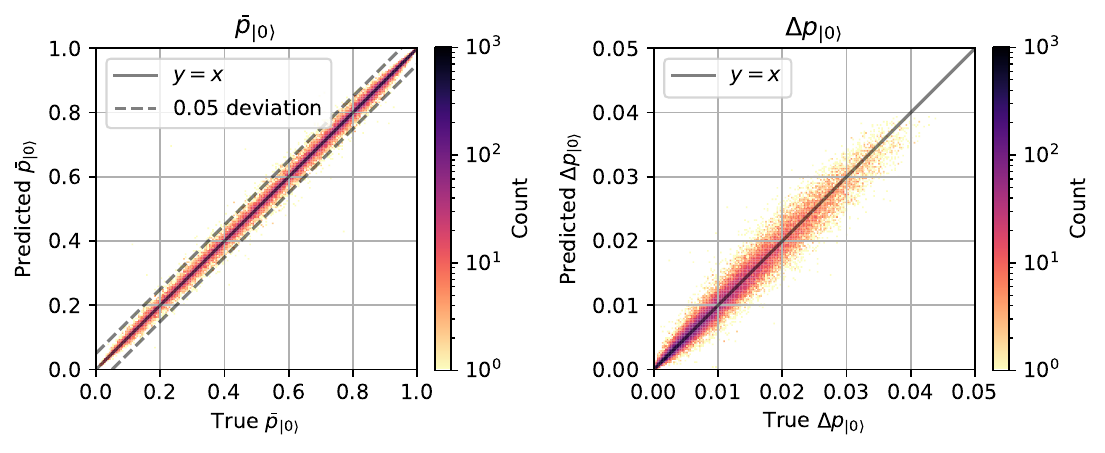}
        \caption{
        True values from the simulation plotted versus the network predictions from the test set. 
        Pulses for which the learned model and the simulation predict the same outcome result in points on the diagonal.
        The dashed lines show the bounds used to calculate the accuracy $\mathcal{A}_{0.05}$.
        }
        \label{plot:nn:simple:correlation_plot}
    \end{figure*}

    The learned model yields the accuracy metrics seen in Tab. \ref{tab:simple:nn_test}, showing that the learned model predicts $\bar{p}_{\ket{0}}$ for more than $99\%$ an error of less then $0.05$.
    The \gls{mae} and \gls{mse} with respect to the two learned objectives are shown in Tab. \ref{tab:simple:nn_test_others}, and the distribution of the predictions with respect to the simulation can be seen in Fig. \ref{plot:nn:simple:correlation_plot}. 
    This shows convergence over the whole $\bar{p}_{\ket{0}}$ range and higher spread towards $\bar{p}_{\ket{0}} = 1/2$, which corresponds to points on the equator of the Bloch sphere. The distribution of the learned standard deviation $\Delta p_{\ket{0}}$ gets wider for larger $\Delta p_{\ket{0}}$, where one contributing factor may be statistical fluctuations from the limited number of shots per probe pulse and a larger intrinsic variation of the outcome in this regime.
    In general Fig. \ref{plot:nn:simple:correlation_plot} shows that the correlation between model prediction and simulation is similar up to small deviations.
    The error with relation to the pulse length can be seen in Fig. \ref{plot:nn:simple:performance_vs_length}, which shows that the error increases only slowly with longer sequences. Hence the learned model is generalizing to sequences longer than what the network has been trained on.

    \begin{table}[]
        \begin{tabular}{l|l|l}
                MAE Loss
                & \multicolumn{1}{c|}{$\mathcal{A}_{0.05}$} 
                & \multicolumn{1}{c}{$\mathcal{A}_{0.01}$} \\ \hline
                \multicolumn{1}{r|}
                {$4.78 \cdot 10^{-3}$} 
                & $99.05\%$
                & $71.25\%$
        \end{tabular}
        \caption{
        Test metrics for network training. The accuracy metrics describe the percentage of test probe pulses where $\overline{p}_{\ket{0}}$ is predicted with an absolute error smaller or equal to $0.05$ or $0.01$.
        The corresponding correlation plots can be seen in Fig. \ref{plot:nn:simple:correlation_plot}.
        The network is trained to minimize the \gls{mae}.
        }
        \label{tab:simple:nn_test}
    \end{table}

    \begin{table}[]
        \begin{tabular}{r|l|l|l}
                \multicolumn{1}{l|}{} & \multicolumn{1}{c|}{$\overline{p}_{\ket{0}}$} & \multicolumn{1}{c|}{$\Delta p_{\ket{0}}$} & \multicolumn{1}{c}{average} \\ \hline
                MAE                                          & $8.62 \cdot 10^{-3}$                 & $9.47 \cdot 10^{-4}$                        & \boldmath$4.78 \cdot 10^{-3}$ \\ \hline
                MSE                                          & $1.9 \cdot 10^{-4}$                 & $2.00 \cdot 10^{-6}$                        & $9.59 \cdot 10^{-5}$         
        \end{tabular}
        \caption{
        Network errors for predicting $\overline{p}_{\ket{0}}$ and $\Delta p_{\ket{0}}$ on the test set, measured by the \gls{mae} ($L^1$) and \gls{mse} ($L^2$) functions. 
        The network is trained to minimize the value printed in bold.
        }
        \label{tab:simple:nn_test_others}
    \end{table}

    \begin{figure}[]
        \includegraphics[width=0.95\linewidth]{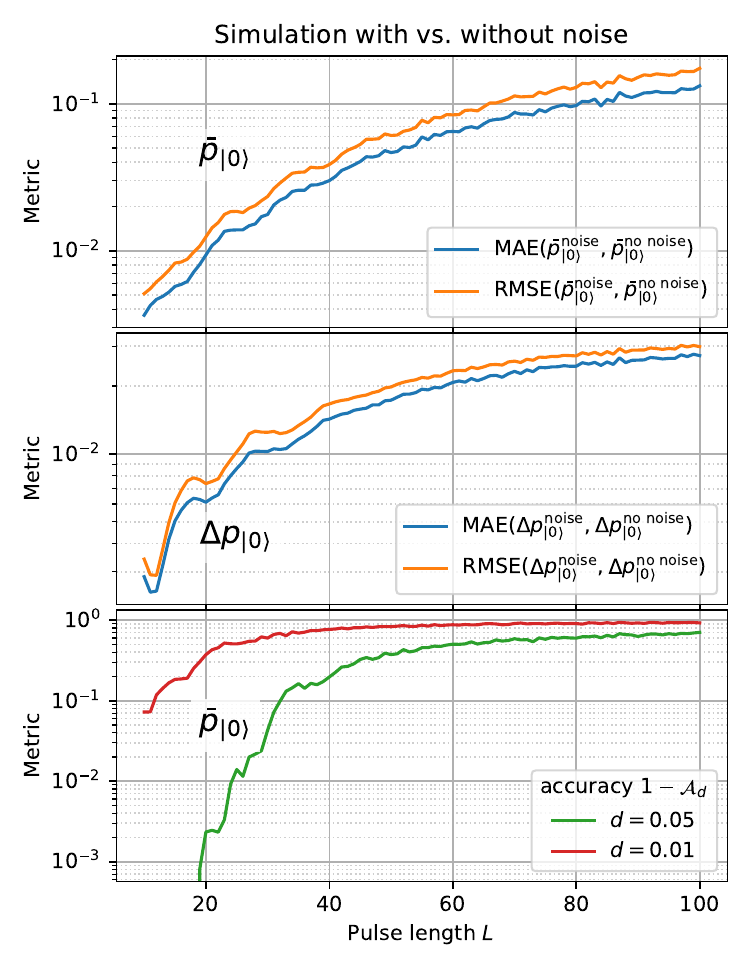}
        \caption{
            Here, we did not train a model, but instead measured the errors introduced by the noise model in the \textit{general} simulation, using the metrics defined to characterize the model’s performance. Specifically, $\overline{p}_{\ket{0}}$ and $\Delta p_{\ket{0}}$ are computed for the test pulses using a Schrödinger equation solver: once without sampling distortions, and once including the noise model via Monte Carlo sampling.
            This plot provides insight into how much the noise alone—independent of the neural network—affects the chosen metrics.
            }
        \label{plot:nn:simple:performance_vs_length_dataset}
    \end{figure}

\subsection{Pulse Optimization}
\label{ssec:Pulse_Optimization}
    After having trained the surrogate model, we now use it to obtain high-fidelity gate sets. We choose to optimize pulses of length $L=\SI{12}{\nano\second}$ with set of \gls{gsc} gate sequences comprised of all permutations of $\{X_{\pi/2}, Y_{\pi/2}\}$ with sequences of up to $4$ consecutive gates.
    The pulses are optimized by means of \gls{sgd}. 
    The used algorithm \ref{alg:opt:general_secription} and the used configurations are stated in the appendix. %
    To finalize the pulse optimization, we choose the $10$ gate sets with the lowest $\mathcal{L}_{GSC}$ measure, resulting in a collection of gate sets with an average infidelity $I = 1.11\%$ and a standard deviation across the selected gate sets of $0.2\%$ (see Fig. \ref{plot:opt:top10} in the appendix).
    From the batch of $256$ gate sets, $46.9\%$ %
    reached an infidelity smaller than or equal to $1\%$. %
    
    The obtained gate infidelities are limited by the intrinsic noise of the simulated qubit device (see Appendix, Fig. \ref{plot:opt:I_coh_vs_I_incoh}). In addition, the initial fidelity used to seed the optimization loop shows no correlation with the final infidelity (Appendix, Fig. \ref{plot:opt:I_initial_vs_I_final}).
    
    To put these values into perspective, we used a conventional pulse optimization (using qopt \cite{qopt}) to optimize $X_{\pi/2}$ and $Y_{\pi/2}$ pulses by directly minimizing the infidelity, without \gls{gsc}, only on the \textit{general} device model. 
    The best $10$ gate sets, out of the $256$ pulses we obtained, reached an average infidelities of $I\textsubscript{sim} = 1.25 \pm 0.04 \%$, When taking the standard deviation over the mean infidelities of the best $10$ gate into account, our method based on the surrogate model and \gls{gsc} performs just as well as an explicit pulse optimization using the simulated model and an accurate fidelity metric (which would both not be readily available in typical experimental settings.)

\subsection{\textit{Specific} Device Simulation}
\label{ssec:Further_Experiments}

    To prove the suitability of this method on a more complex system we also implemented a simulation based on measurements from real-world devices. Hence, we test our method on this \textit{specific} model. 

    As the \textit{specific} transfer function introduces transients, we wait a few segments after the control pulse while applying a small constant control voltage. This ensures that the pulse decayed before the next pulse is applied. Specifically, the voltage $\epsilon_h = \SI{-3.18}{\milli\volt}$ is held for $4$ segments. %
    While optimizing pulses, this is enforced by fixing these 4 pulse segments to $\epsilon_h$.
    Additionally, $\SI{6}{\nano\second}$ at $\epsilon_h$ are added to the end of the entire pulse sequence before the qubit state is measured.
    
    The size of the network has been adapted to this more complex \textit{specific} simulation by increasing the number of neurons per layer and omitting the batch normalization layers. The difference between the networks is described in the appendix.
    To adapt the device sampling process to the markedly different exchange interaction of the \textit{specific} model (see Fig. \ref{plot:exchange_interactions}), we reduce the relative number of probe pulses based on the assumed exchange interaction.

    To obtain a lower bound, we choose the system realization and the voltage range, to be more difficult than one would expect to encounter when applying our protocol to an experiment. In practice, the operation regime can be chosen such that the deviations from common approximations are smaller. 

\begin{figure}[!h]
        \includegraphics[width=0.95\linewidth]{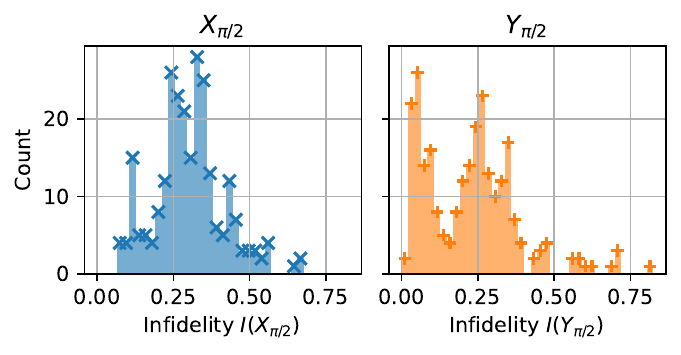}
        \caption{
            Distribution of obtained final incoherent infidelities of the $256$ optimized gate sets for the \textit{specific} simulation.
        }
        \label{plot:opt:v10_hists_moresyse}
    \end{figure}

    Nonetheless, the trained model was able to predict the measurement outcome of $95.28\%$ of the test pulses simulated using the \textit{specific} model with an absolute error $\leq 0.05$. The average absolute deviation from the correct outcome of $\overline{p}_{\ket{0}}$ was $1.66 \cdot 10^{-2}$ (see Tab. \ref{tab:numeric_model:final_test_metrics} and Tab. \ref{tab:numeric_model:final_test_metrics_other}). 
    As Tab. \ref{tab:numeric_model:final_data}, Fig. \ref{plot:opt:more_syse_I_scatter_unsorted}, and Fig. \ref{plot:opt:v10_hists_moresyse} show, the obtained gate set infidelities scatter, but $I<20\%$ is reached frequently. Thus our method reaches the needed infidelities to initialize the \gls{gsc} protocol for this system \cite{Cerfontaine_2020}, even in this worst-case scenario.

    \begin{table}[]
        \begin{tabular}{l|l|l}
            MAE Loss                                & 
            \multicolumn{1}{c|}{$\mathcal{A}_{0.05}$} & 
            \multicolumn{1}{c}{$\mathcal{A}_{0.01}$} \\ \hline
            \multicolumn{1}{r|}{
            $9.21 \cdot 10^{-3}$} & 
            $95.28\%$    &
            $44.41\%$                                
        \end{tabular}
        \caption{
            Test metrics for network training for the \textit{specific} system. The accuracy $\mathcal{A}_d$ metrics describe the percentage of test points where $\overline{p}_{\ket{0}}$ is predicted with an error smaller or equal to $d$.
        }
        \label{tab:numeric_model:final_test_metrics}
    \end{table}
    
    \begin{table}[]
        \begin{tabular}{r|l|l|l}
            \multicolumn{1}{l|}{} & 
            \multicolumn{1}{c|}{$\overline{p}_{\ket{0}}$} & 
            \multicolumn{1}{c|}{$\Delta p_{\ket{0}}$} & 
            \multicolumn{1}{c}{average} \\ \hline
            $L^1$                                          &
            $1.66 \cdot 10^{-2}$                 &
            $1.78 \cdot 10^{-3}$                        & 
            \boldmath$9.21 \cdot 10^{-3}$ \\ \hline
            $L^2$                                          & 
            $5.46 \cdot 10^{-4}$                 & 
            $5.87\cdot 10^{-6}$                        & 
            $2.76 \cdot 10^{-4}$         
        \end{tabular}
        \caption{
            Network errors for predicting $\overline{p}_{\ket{0}}$ and $\Delta p_{\ket{0}}$ on the test set sampled from the simulation using the \textit{specific} model. 
            Network is trained to minimize value printed in bold. 
            (measured by the \gls{mae} ($L^1$) and \gls{mse} ($L^2$) functions)
        }
        \label{tab:numeric_model:final_test_metrics_other}
    \end{table}
    
    \begin{table}[]
        \begin{tabular}{r|l}
            \multicolumn{1}{l|}{Average Infidelity of 256 gate sets} & 
            \multicolumn{1}{c}{$ 26.12 \% $} \\ \hline %
            \multicolumn{1}{l|}{Standard deviation of 256 gate set infidelities} & 
            \multicolumn{1}{c}{$ 9.60 \% $}\\ \hline %
            \multicolumn{1}{l|}{Best infidelity of one gate set} & 
            \multicolumn{1}{c}{$ 5.24\% $} %
        \end{tabular}
        \caption{
            Resulting infidelities from optimization using the \textit{specific} model obtained using our method and minor adjustments compared to the application on the \textit{general} simulation. The performance can be further improved by optimizing the sampling method and \gls{nn} parameters.
        }
        \label{tab:numeric_model:final_data}
    \end{table}

\section{Conclusion and Outlook}
\label{sec:Conclusion_and_Outlook}
    
    In this work, we present a fully data-driven approach for obtaining self-consistent, high-fidelity gate sets from minimal prior information. A surrogate model trained on easily accessible experimental data is shown to successfully guide numerical pulse optimization. Strikingly, our method achieves performance comparable to optimization schemes that rely on full knowledge of the simulation model, despite not having such access. For complex and realistic device dynamics (captured by our \textit{specific} model), the resulting infidelities after pulse optimization achieve fidelities required for applying the \gls{gsc} protocol.
    \\
    Future directions include implementing this protocol on an experimental platform, examining the extent to which the resulting gates display dynamically error-correcting behavior, extending the approach to other qubit types and shuttling dynamics, and investigating scalability to multi-qubit systems. Additional promising avenues involve leveraging online learning to efficiently estimate gradients during optimal control or adaptive parameter-space exploration. It will also be valuable to assess how different probe pulse-design strategies scale with mismatches between assumed models and actual devices.

\section{Data Availability}
    The Python code and data sets necessary to reproduce the main results of this study are publicly available in the Zenodo repository at https://doi.org/10.5281/zenodo.18354586.

\begin{acknowledgments}
    This work was funded by the German Federal Ministry of Education and Research, reference numbers 13N15652 and by the Deutsche Forschungsgemeinschaft (DFG, German Research Foundation) under Germany's Excellence Strategy – Cluster of Excellence Matter and Light for Quantum Computing (ML4Q) EXC 2004/1 – 390534769.

    We thank Felix Motzoi for the inspiring discussions at the beginning of this project. In addition, we would like to thank Christian Gorjaew for providing us with the fit to the exchange interaction measurement.
    
\end{acknowledgments}

\appendix

    \section{Further Details About The Network Training}
    \label{sec:nn:architecture_summary}
    
    The \gls{rnn} is trained using the \gls{mae} loss function:
    \begin{equation}
        \mathcal{L}_{MAE} = \frac{1}{N} \sum_i \abs{t_i-y_i},
    \end{equation}
    $t_i$ is the target outcome for data point $i$ and $y_i$ is the current network prediction of the outputs $\overline{p}_{\ket{0}}$ and $\Delta p_{\ket{0}}$.
    Further, the loss function \gls{rmse}, used for the discussions in this paper, is defined by
    \begin{equation}
        \mathcal{L}_{RMSE} = \sqrt{\frac{1}{N} \sum_i \abs{(t_i-y_i)}^2}.
    \end{equation}

    Unlike \gls{mse}, \gls{mae} penalizes outliers less strongly, making it more robust in noisy settings.
    It has been shown that $L^1$ leads to less pronounced noisy artefacts in comparison to $L^2$-based loss functions in image restoration tasks \cite{zhao2018loss}. 
    But for this task, the \gls{mae} loss is not essential: The mean square error can also be used for the loss function for this kind of task \cite{Krastanov_2019}, as well as the cross-entropy loss function \cite{Flurin_2020}.
    
    To address the imbalance between the target quantities, additional scaling can be applied during training. Specifically, $\overline{p}_{\ket{0}}$ and $\Delta p_{\ket{0}}$ are not naturally of the same order of magnitude, which could be equalized by scaling their contribution to the loss function. We did not scale their contributions in this work. 
    
    To compensate uneven sampling on the Bloch sphere, the contribution of each data point to the \gls{mae} loss is weighted according to the sampling density along the $\overline{p}_{\ket{0}}$ axis. Concretely, the distribution of $\overline{p}{\ket{0}}$ is partitioned into bins of width $0.01$. For each bin, the number of samples is compared to the count of the most populated bin. A scaling factor is then derived by normalizing against this maximum count, and the factor is multiplied with the training error of each data point belonging to that bin. In this way, regions of the $\overline{p}{\ket{0}}$ space with lower data density are upweighted, ensuring that the network does not neglect them during optimization.

    For the optimizer, we used Adam \cite{kingma2017adam}, starting with a learning rate of $lr = 0.001$, which was reduced by a factor of $0.8$ after every $60$ epochs without improvement on the validation set. Training was stopped if no improvement occurred for $130$ consecutive epochs. For the \textit{specific} simulation, the learning rate was reduced by a factor of $0.9$ after $25$ epochs without improvement, and training was stopped after $60$ epochs without improvements on the validation set.

    The network for the \textit{general} simulation was trained on $2,000,000$ samples with pulse durations $T \in [\SI{10}{\nano\second}, \SI{50}{\nano\second}]$. The dataset was split into training, validation, and test sets, consisting of $81\%$, $9\%$, and $10\%$ of the samples, respectively.
    
    The network used for the \textit{general} simulation is structured as follows. 
    The input pulse is encoded into a tensor of shape (None, $L_{\max}$, 3), where $L_{\max}$ is the maximum pulse length the network is expected to handle. The three channels correspond to $\epsilon_t$, $\Delta B_Z$, and $dt$, with $dt \in \{0,1\}$ serving as a mask for values acting on the device (None denotes the variable batch dimension). 
    For shorter sequences with $L < L_{\max}$, the remaining values are padded with zeros.
    The pulse is first processed by convolutional layers with kernel width $3$, resulting in a tensor of shape (None, $L_{\max}$, 64). These $3 \times 3$ convolutional layers span three time steps, allowing local temporal information to be captured. This encoding is then refined by $1 \times 1$ convolutions mapping to (None, $L_{\max}$, 20), with batch normalization applied between layers.
    The resulting (None, $L_{\max}$, 20) representation is passed through a standard \gls{lstm} layer iterating along the time dimension of length $L_{\max}$, yielding an output of shape (None, 100). The final dense layers, fed with the last hidden state of the \gls{lstm}, map this representation to the predicted outputs $\overline{p}_{\ket{0}}(\boldsymbol{\lambda})$ and $\Delta p_{\ket{0}}(\boldsymbol{\lambda})$ with shape (None, 2). The outputs are clipped to $[0, 1]$ to ensure interpretability as probabilities.

    \begin{table}[]
        \centering
        \begin{tabular}{|c|c|}
            \hline
            Type & Configuration \\
            \hline
        Zero Padding & 1 segment before and after input control pulse \\
            1D Conv & 3 wide kernel; (16, 32, BN, 64); selu activation \\
            1D Conv & 1 wide kernel; (70, BN, 50, 20); sin activation \\
            LSTM & 100 units; default tensorflow implementation \\
            Dense layers & (100, 70, 10, 2); relu activation \\
            Clipping & Two output nodes with values clipped to $[0, 1]$\\
            \hline
        \end{tabular}
        \caption{
            Network architecture of the \textit{general} system, containing $83,562$ trainable parameters. Values in brackets denote the number of kernels in convolutional layers or the number of neurons in dense layers. BN indicates a Batch Normalization layer.
        }
        \label{tab:nn:architecture_summary}
    \end{table}
    
    The model used for the \textit{specific} simulation uses more neurons per layer and can be summarized as shown in Tab. \ref{tab:nn:architecture_summary_moresyse}.
    \begin{table}[]
        \centering
        \begin{tabular}{|c|c|}
            \hline
            Type & Configuration \\
            \hline
            Zero Padding & 1 segment before and after input control pulse \\
            1D Conv & 3 wide kernel; (30, 50, 80); selu activation \\
            1D Conv & 1 wide kernel; (90, 80, 50); sin activation \\
            LSTM & 150 units; default tensorflow implementation \\
            Dense layers & (150, 100, 20, 2); relu activation \\
            Clipping & Two output notes with values clipped to $[0, 1]$\\
            \hline
        \end{tabular}
        \caption{
            Network architecture of the \textit{specific} system, containing $195,962$ trainable parameters. Values in brackets denote the number of kernels in convolutional layers or neurons in dense layers. BN indicates a Batch Normalization layer.
        }
        \label{tab:nn:architecture_summary_moresyse}
    \end{table}

    \section{Further Device Simulation Information}

    Note that we did not use the exact notation of Eq.~\ref{eq:main_H} in the simulation, but instead added the noise directly into $\boldsymbol{\lambda}$. By tailoring the exponential function introduced in Eq.~\ref{eq:H_c}, which is used to obtain $\boldsymbol{\lambda}$, one recovers a Hamiltonian of the form given in Eq.~\ref{eq:main_H} \cite{hangleiter2021filter}.

    The exchange interaction function for the \textit{specific} device, as shown in Fig.~\ref{plot:exchange_interactions}, which has been obtained from \cite{PhysRevLett.110.146804}, can be written as 
    \begin{equation}
        J(\epsilon) = J_0 e^{\epsilon/\epsilon_0} - \frac{1}{2}\left(1+\tanh{\frac{\epsilon-\epsilon_s}{w}}\right)\left(J_0e^{\epsilon/\epsilon_0}-\left(b+m\epsilon\right)\right)
        \label{eq:numeric_exchange_interaction_fit_function}
    \end{equation}
    
    \begin{table}[]
        \centering
        \begin{tabular}{c|c}
                fit parameter & value \\ \hline
                $J_0$ & $\SI{0.367}{\giga\hertz}$\\
                $\epsilon_0$ & $\SI{0.79}{\milli\volt}$\\
                $\epsilon_s $ & $\SI{0.21}{\milli\volt}$\\
                $w$ & $\SI{0.461}{\milli\volt}$\\
                $b$ & $\SI{22.720}{\giga\hertz}$\\
                $m$ & $\SI{1.892}{\giga\hertz\per\milli\volt}$
        \end{tabular}
        \caption{
        Fit parameters for Eq.~\ref{eq:numeric_exchange_interaction_fit_function}, with the resulting function shown in orange in Fig.~\ref{plot:exchange_interactions}.
        }
        \label{tab:J_fit_params}
    \end{table}

\section{Details About The Pulse Optimization}
\label{ssec:appendix_details_about_the_pulse_optimization}
            
    \begin{algorithm}[]
            choose number of gate sets $N$ (default $256$)\;
            choose number of epochs $N_e$\;
            initialize $\boldsymbol{\lambda}_n(G) : \forall n \in [0, N-1], G \in \mathcal{G}$\;
            choose $\Delta p_{\ket{0}}$ regularization $\gamma$ (default $1$)\;
            choose mini-batch size $K_b$ with $N \mod K_b = 0$ (default $1$)\;
            choose mixing factor $\delta$ (default $0$)\;
            
            \For{$n_e \in [0, N_e]$}{
                    \For{$b \in [0, N/K_b-1]$}{
                            \For{$n \in [b K_b, (b+1) K_b]$}{
                                    $\forall S_i \in \mathcal{S} : \mathcal{R}_{n, i} = \overline{p}_{\ket{0}}(\boldsymbol{\lambda}_n(S_i))$\;
                                    $\mathcal{L}_{GSC, n} = \frac{1}{N_{seq}} \sum_{S_i \in \mathcal{S}} |\mathcal{R}_{n, i} - R_i|^2$\;
                                    $\mathcal{L}_{std, n} = \frac{1}{N_s} \sum_{S_i \in \mathcal{S}} |\Delta p_{\ket{0}}(\boldsymbol{\lambda}_n(S_i))|^2$\;
                                    $\mathcal{L}_n = \mathcal{L}_{GSC, n} + \gamma \mathcal{L}_{std, n}$\;
                                    $\forall G \in \mathcal{G} : g_{n, G} = \frac{\partial \mathcal{L}_n}{\partial \boldsymbol{\lambda}_n(G)}$\;
                            }

                            $\forall G \in \mathcal{G} : \bar{g}_G = \frac{1}{K_b} \sum_n g_{n, G}$\;
                            $\forall G \in \mathcal{G} : g_{n, G} = (1-\delta) g_{n, G} + \delta \bar{g}_G$\;
                            
                            request optimizer (e.g. SGD or Adam) to update $\boldsymbol{\lambda}_n(G)$ using $g_{n, G}$ : $\forall G \in \mathcal{G}$\;
                    }
            }
            \label{alg:opt:general_secription}
            \caption{Pulse optimization loop including mini-batching and $\Delta p_{\ket{0}}$ regularization.}

    \end{algorithm}

    Depending on the initialization of the control parameters $\boldsymbol{\lambda}$, the optimization of $\mathcal{L}_{GSC}$ occasionally converged to local minima where not all syndromes matched. To mitigate this, we began with shorter sequences of at most three gates and temporarily increased the exponent in $\mathcal{L}_{GSC}$ (Eq.~\ref{eq:L_GSC}) from 2 to 4 during the initial iterations. Incorporating $\mathcal{L}_{std}$ is not strictly required, but serves to suppress statistical noise in the pulses. Preliminary tests suggest that for a strong influence of this regularization term, a value of $\gamma \approx 10$ is appropriate; further work is needed to confirm whether this indeed reduces statistical errors. Finally, we did not explore gradient averaging across batch elements and therefore set $\delta = 0$, which we consider the most consistent choice.

    We have chosen \gls{sgd} for the pulse optimization phase. To accelerate convergence, more advanced optimizers such as Levenberg–Marquardt or quasi-Newton methods like L-BFGS could be considered.
    
    We executed the optimization loop described in Algorithm~\ref{alg:opt:general_secription} using the configurations listed in Tab.~\ref{tab:opt:optimization_steps}.

    \begin{table}[]
        \centering
        \begin{tabular}{|c|c|c|}
        \hline
        Iterations & SGD $l_r$ & Configuration \\
        \hline
            $6000$ & 0.5 & $L_S = 3$, $\mathcal{L}_{GSC}$ exponent 2, $K_b = 1$, $\gamma=0$ \\
            $6000$ & 0.5 & $L_S = 4$, $\mathcal{L}_{GSC}$ exponent 4, $K_b = 1$ \\
            $22500$ & 0.05 & $\mathcal{L}_{GSC}$ exponent 2, $K_b = 8$, $\delta = 0.05$ \\
            $10000$ & 0.03 &  \\
            $22500$ & 0.04 & $\gamma=2$ \\
            $10000$ & 0.03 & \\
            \hline
        \end{tabular}
        \caption{
            Learning rates $l_R$ for \gls{sgd} and related configurations used to optimize pulses of the \textit{general} model in Algorithm~\ref{alg:opt:general_secription}. 
            Parameters given for later iterations update those from previous ones. 
            Changes in the maximal sequence length $L_S$ of the \gls{gsc} sequences are shown as the dashed line in Fig.~\ref{plot:opt:simple:L_S_vs_epoch}.
        }
        \label{tab:opt:optimization_steps}
    \end{table}

\section{Global Z-rotation consideration}
\label{ssec:appendix_gloabl_z_rotaion_consideration}
    
    To calculate the fidelities of the obtained gate sets $\{X'_{\pi/2}, Y'_{\pi/2}\}$ with respect to the self-consistent Z-rotation, we first obtain the total evolution from the used qopt \cite{qopt} simulation package, convert this to a rotation matrix, and then decompose it into ZXZ Euler angles.
    \begin{equation}
        X'_{\pi/2} = R_Z(\varphi_{X, Z1}-\theta) \cdot R_X(\varphi_{X, X1}) \cdot R_Z(\varphi_{X, Z2}+\theta)
    \end{equation}
    \begin{equation}
        Y'_{\pi/2} = R_Z(\varphi_{Y, Z1}-\theta) \cdot R_X(\varphi_{Y, X1}) \cdot R_Z(\varphi_{Y, Z2}+\theta).
    \end{equation}
    Then we solve for $\theta$ with the expected ideal values for $\varphi_{*, Z1}$ and $\varphi_{*, Z2}$.
    From there we correct for the global z rotation via
    \begin{equation}
        X''_{\pi/2} = R_Z(+\theta) \cdot X'_{\pi/2} \cdot R_Z(-\theta)
    \end{equation}
    \begin{equation}
        Y''_{\pi/2} = R_Z(+\theta) \cdot Y'_{\pi/2} \cdot R_Z(-\theta)
    \end{equation}
    The fidelities are then calculated based on $\{X''_{\pi/2}, Y''_{\pi/2}\}$ using the entanglement fidelity measure
    \begin{equation}
        F = 1-I = \frac{1}{d^2} \abs{Tr(U^\dagger V)}^2,
    \end{equation}
    where $U$ is the target operator, $V$ the realized operator, $d$ is the dimensionality of the Hilbert space, and $I$ is the infidelity.
    
    \begin{figure}[!h]
            \includegraphics[width=0.95\linewidth]{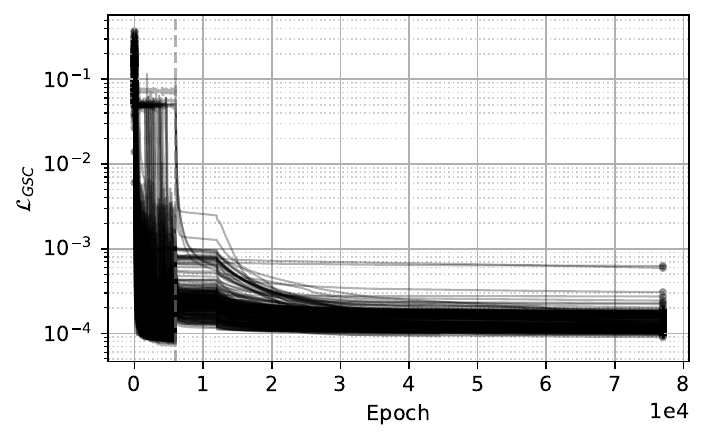}
            \caption{
                The history of $\mathcal{L}_{GSC}$ while optimizing pulses for the \textit{general} model.
            }
            \label{plot:opt:simple:L_S_vs_epoch}
    \end{figure}

    \begin{figure}[!h]
            \includegraphics[width=0.95\linewidth]{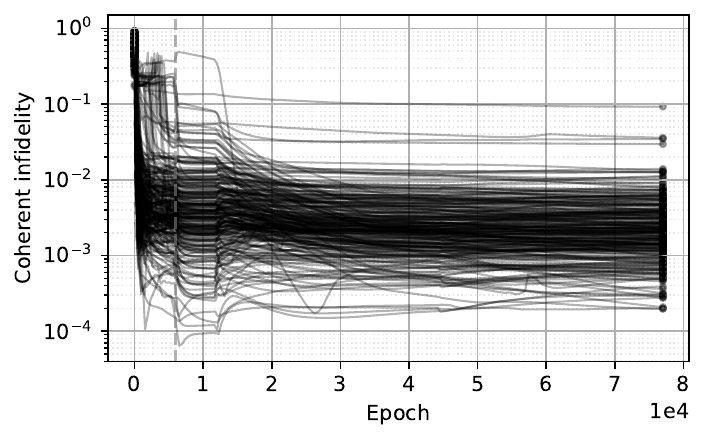}
            \caption{
                The history of infidelity calculated without noise $I_{coh}$ while optimizing for the \textit{general} model.
            }
            \label{plot:opt:simple:L_coh_vs_epoch}
    \end{figure}

    \begin{figure}[!h]
            \includegraphics[width=0.95\linewidth]{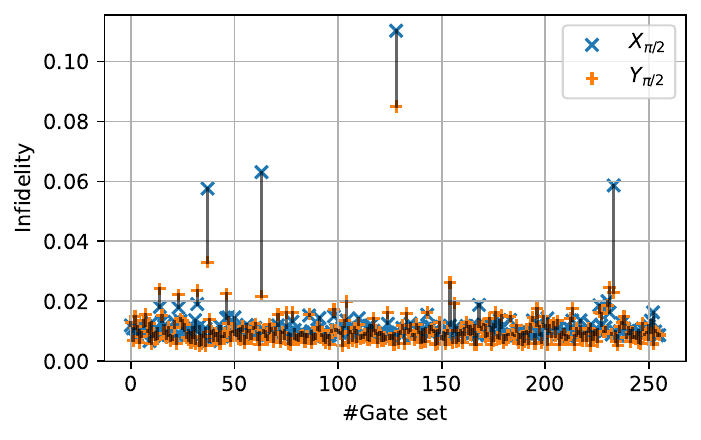}
            \caption{
                Distribution of incoherent infidelity measured with noise of $256$ gate sets obtained from optimizing $\mathcal{L}_{GSC}$ using the learned qubit model for the \textit{general} simulation.
                The black lines visualize the infidelity difference between the two gates of one self-consistent gate set.
            }
            \label{plot:opt:I_by_gate_set_number}
    \end{figure}

    \begin{figure}[!h]
            \includegraphics[width=0.95\linewidth]{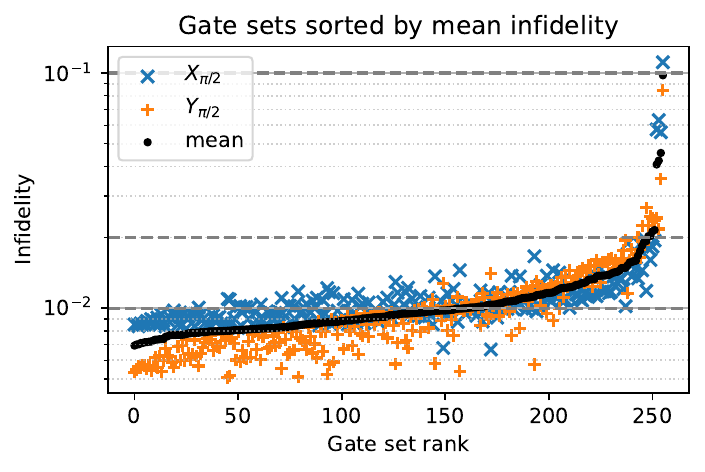}
            \caption{
            Gate sets sorted by their infidelity measured with noise. The $Y_{\pi/2}$ gates reach smaller infidelities which we assume to be the case due to the starting conditions and the control parameter in the Hamiltonian, the path length of the trajectory for realizing a $Y_{\pi/2}$ rotation is shorter, than for a $X_{\pi/2}$ rotations.
            }
            \label{plot:opt:I_sorted}
    \end{figure}

    \begin{figure}[!h]
            \includegraphics[width=0.95\linewidth]{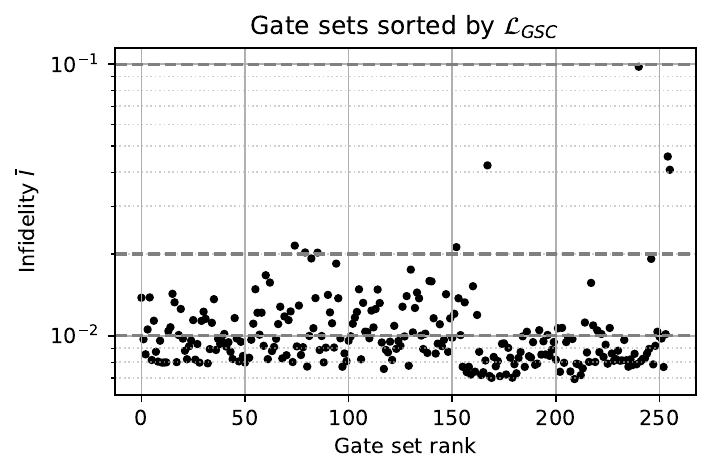}
            \caption{
            Infidelities of gate sets sorted by $\mathcal{L}_{GSC}$. Outliers, where the optimization did not converge successfully, tend to have higher $\mathcal{L}_{GSC}$.
            }
            \label{plot:opt:I_sorted_by_L_S}
    \end{figure}
    
    \begin{figure}[!h]
            \includegraphics[width=0.95\linewidth]{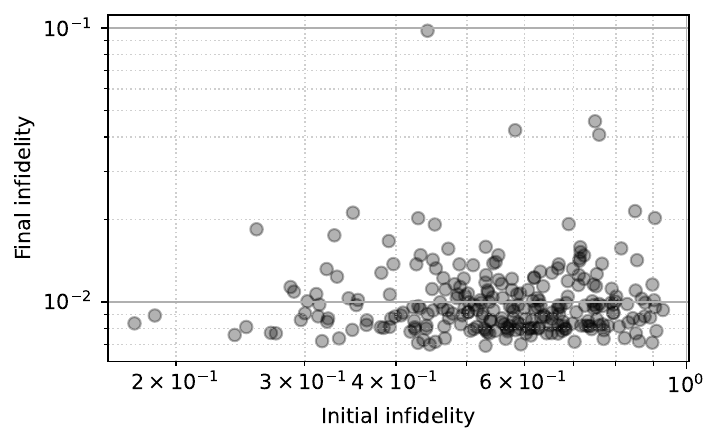}
            \caption{
            Scatter plot visualizing the independence of the final optimized infidelities on the initial infidelities for the \textit{general} simulation.
            }
            \label{plot:opt:I_initial_vs_I_final}
    \end{figure}
    
    \begin{figure}[!h]
            \includegraphics[width=0.95\linewidth]{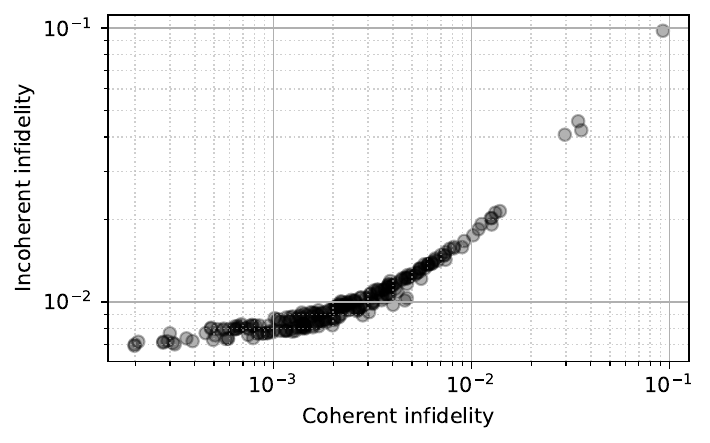}
            \caption{
            Scatter plot showing the correlation between incoherent and coherent infidelities of the final gate sets for the \textit{general} simulation. 
            For smaller coherent infidelities, the incoherent infidelities converge to approximately $8 \cdot 10^{-3}$ %
            and stochastic errors start to become limiting. 
            }
            \label{plot:opt:I_coh_vs_I_incoh}
    \end{figure}

    \begin{figure}[!h]
            \includegraphics[width=0.95\linewidth]{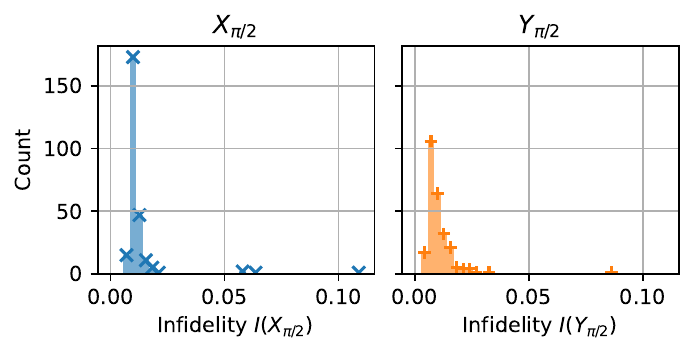}
            \caption{
                Distribution of obtained final incoherent infidelities of the 256 optimized gate sets for the \textit{general} simulation.
            }
            \label{plot:opt:v10_hists}
    \end{figure}
    
    \begin{figure}[]
            \includegraphics[width=\linewidth]{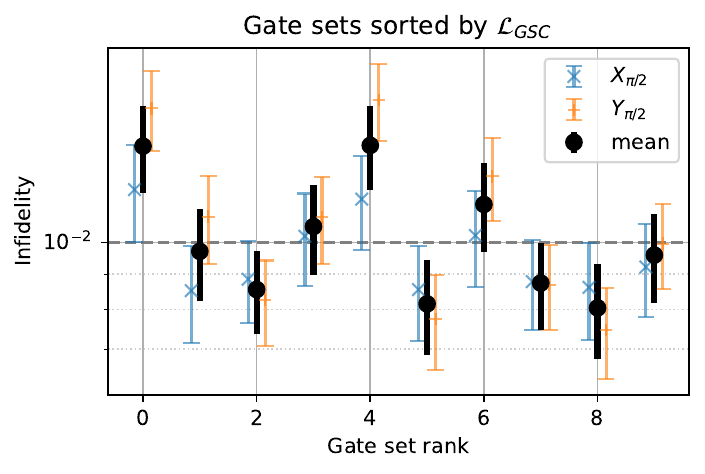}
            \caption{
            The ten best gate sets selected by $\mathcal{L}_{GSC}$. The orange and blue bars represent the mean infidelities of $X_{\pi/2}$ and $Y_{\pi/2}$, respectively, each averaged over 60 noise samples with error bars denoting the standard error. The black bar labeled "mean" indicates the overall average of the orange and blue values. Each of these gate sets was then used as an initial guess for running \gls{gsc} on the experiment.
            }
            \label{plot:opt:top10}
    \end{figure}
    
    \section{Details To The Pulse Optimization For The \textit{Specific} Device Simulation}
\begin{figure}[!h]
            \includegraphics[width=0.95\linewidth]{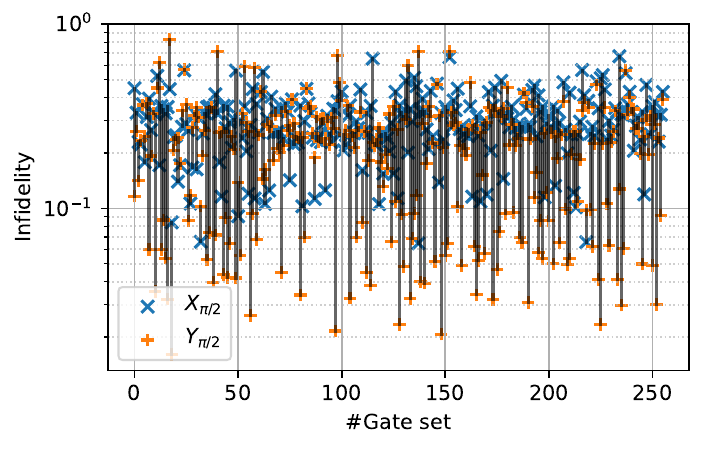}
            \caption{
                The infidelities of the obtained gates from the \textit{specific} simulation. 
                The black lines connect the $X_{\pi/2}$ and $Y_{\pi/2}$ marker of one gate set.
            }
            \label{plot:opt:more_syse_I_scatter_unsorted}
    \end{figure}

    The distribution of the obtained gate set fidelities can be seen in Fig. \ref{plot:opt:more_syse_I_scatter_unsorted}. The corresponding histogram is shown in Fig. \ref{plot:opt:v10_hists_moresyse}.
        
    \begin{figure}[!h]
            \includegraphics[width=0.95\linewidth]{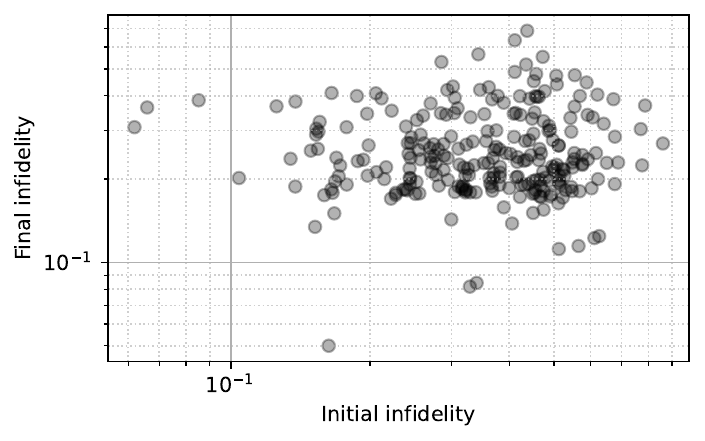}
            \caption{
                For the \textit{specific} model: Scatter plot visualizing the independence of the final infidelities on the initial infidelities with which the optimization is run for the \textit{specific} simulation.
            }
            \label{plot:opt:initial_final_infidelity_scatter_moresyse}
    \end{figure}
    \FloatBarrier

\bibliography{main.bib} %

\end{document}